\documentclass[%
5p,
times,
sort&compress
]{elsarticle}

\usepackage[T1]{fontenc}

\usepackage[english]{babel}

\usepackage[hidelinks]{hyperref}

\usepackage{mathtools} 
\usepackage{amsfonts}
\allowdisplaybreaks%

\usepackage{tensor}
\newcommand{\ind}{\indices}

\def\mf{\mathfrak}
\def\pd{\partial}
\def\mring{\mathring}
\def\mc{\mathcal}
\def\mrm{\mathrm}

\begin{document}

\title{Dark energy as a kinematic effect}

\author[ift]{H.~Jennen\corref{cor}}
\ead{hjennen@ift.unesp.br}

\author[ift]{J.~G.~Pereira}
\ead{jpereira@ift.unesp.br}

\cortext[cor]{Corresponding author}

\address[ift]{%
  Instituto de F\'{\i}sica Te\'orica, UNESP - Universidade 
  Estadual Paulista, \\
  Rua Dr.~Bento Teobaldo Ferraz, 271 -- Bl.~II, 01140-070, S\~ao 
  Paulo, SP,  Brazil
}
\date{\today}

\begin{abstract}
We present a generalization of teleparallel gravity that is 
consistent with local spacetime kinematics regulated by the de 
Sitter group $SO(1,4)$. The mathematical structure of 
teleparallel gravity is shown to be given by a nonlinear 
Riemann--Cartan geometry without curvature, which inspires us to 
build the generalization on top of a de Sitter--Cartan geometry 
with a cosmological function. The cosmological function is given 
its own dynamics and naturally emerges nonminimally coupled to 
the gravitational field in a manner akin to teleparallel dark 
energy models or scalar-tensor theories in general relativity.  
New in the theory here presented, the cosmological function gives 
rise to a kinematic contribution in the deviation equation for 
the world lines of adjacent free-falling particles. While having 
its own dynamics, dark energy manifests itself in the local 
kinematics of spacetime.
\end{abstract}


\maketitle

\section{Introduction}

Physically equivalent to general relativity in its description of 
the gravitational interaction, teleparallel gravity is 
mathematically and conceptually rather different from Einstein's 
\emph{opus magnum}. Although the precise implementation of 
general relativity differs from the one of teleparallel gravity, 
their geometric structures are related by switching between 
certain subclasses of Riemann--Cartan spacetimes. On the one 
hand, general relativity being the standard model for classical 
gravity, it is naturally well known that the fundamental field is 
the vierbein, which is accompanied by the Levi-Civita spin 
connection. The resulting spacetime is thus characterized by 
a Riemann--Cartan geometry without torsion. On the other hand, 
teleparallel gravity takes a different route in order to 
generalize the geometry of Minkowski space to incorporate the 
dynamics of the gravitational field, for it is torsion instead of 
curvature that is turned on by gravitating 
sources~\cite{aldrovandi:2012tele}. Since the curvature remains 
zero, the spin connection retains its role of representing 
inertial effects only. Therefore, the description can be extended 
to account for a breakdown of the weak equivalence 
principle~\cite{Aldrovandi:2003pa}.

The rationale behind Riemann--Cartan geometry underlying both 
theories is closely related to the assumption that kinematics is 
locally governed by the Poincar\'e group $ISO(1,3)$. Be that as 
it may, there is significant evidence that our universe 
momentarily undergoes accelerated expansion~\cite{Peebles:2003cc, 
  Weinberg:2008bc}, which indicates that the large-scale 
kinematics of spacetime is approximated better by the de Sitter 
group $SO(1,4)$~\cite{Aldrovandi:2006vr}. We shall take this 
evidence to heart and conjecture that local kinematics is 
regulated by the de Sitter group. Looked at from a mathematical 
standpoint, this amounts to have the Riemann--Cartan geometry 
replaced by a Cartan geometry modeled on de Sitter 
space~\cite{Wise:2010sm}. The corresponding spacetime is 
everywhere approximated by de Sitter spaces, whose combined set 
of cosmological constants in general varies from event to event, 
hence resulting in the cosmological 
function~\cite{Jennen:2014mba}.

In the present article we propose an extended theory of gravity 
as we generalize teleparallel gravity for such a de 
Sitter--Cartan geometry. Quite similar to the cosmological 
constant in teleparallel gravity or general relativity, we model 
the dark energy driving the accelerated expansion by 
a cosmological function $\Lambda$ of dimension one over length 
squared. Fundamentally different, however, the cosmological 
function alters the kinematics governing physics around any 
point, such that spacetime is approximated locally by a de Sitter 
space of cosmological constant $\Lambda$. To be exact, 
a congruence of particles freely falling in an external 
gravitational field exhibits a relative acceleration, not only 
due to the nonhomogeneity of the gravitational field, but also 
because of the local kinematic properties of spacetime that are 
determined by the cosmological function.

The organization of the article is as follows. In 
Sec.~\ref{sec:dSCgeo} the basic tools for de Sitter--Cartan 
geometry that are used in subsequent sections are reviewed 
briefly. Afterwards, we show in Sec.~\ref{sec:tg} that the 
mathematical structure underlying teleparallel gravity is that of 
a nonlinear Riemann--Cartan geometry, which is consistent with 
the standard interpretation of teleparallel gravity as a gauge 
theory for the Poincar\'e translations. This is an important 
observation, since we leave the gauge picture for what it is upon 
generalizing the theory for a cosmological function in de Sitter 
teleparallel gravity. Section~\ref{sec:dStg} is devoted to the 
main results of this article, in which we discuss in sequence the 
fundamentals of de Sitter teleparallel gravity, the phenomenology 
of the kinematic effects, and the dynamics of the gravitational 
field and the cosmological function. We conclude 
in~Sec.~\ref{sec:concl}.

\section{de Sitter--Cartan geometry with a cosmological function}
\label{sec:dSCgeo}

The Cartan geometry modeled on $(\mf{so}(1,4),SO(1,3))$ 
consists of a principal Lorentz bundle over spacetime together 
with an $\mf{so}(1,4)$-valued one-form, which is called the 
Cartan connection~\cite{Alekseevsky:1995cc, sharpe1997diff_geo, 
  Wise:2009fu, Wise:2010sm}. The Lie algebra $\mf{so}(1,4)$ that 
generates the de Sitter group is subject to the commutation 
relations
\begin{align}
  -i[M_{ab},M_{cd}] &= \eta_{ac}M_{bd} - \eta_{ad}M_{bc} 
  + \eta_{bd}M_{ac} - \eta_{bc}M_{ad},
  \notag\\
  -i[M_{ab},P_c] &= \eta_{ac}P_b- \eta_{bc}P_a,
  \notag\\
  -i[P_a,P_b] &= -l^{-2}M_{ab},
  \label{eq:comm_relations_so(1,4)}
\end{align}
where $\eta_{ab} = (+, - , -, -)$ and we parametrize elements of 
$\mf{so}(1,4)$ by $\tfrac{i}{2} \lambda^{ab} M_{ab} + i \lambda^a 
P_a$, so that the $M_{ab}$ span the Lorentz subalgebra, while the 
$P_a \equiv M_{a4}/l$ are a basis for the de Sitter 
transvections. The length scale $l$ is related to the 
cosmological constant of the corresponding de Sitter space 
$SO(1,4)/SO(1,3)$, namely,~\cite{Wise:2010sm}
\begin{equation}
  \label{eq:rel.Ll}
  \Lambda = \frac{3}{l^2}.
\end{equation}

Because the Cartan connection is valued pointwise in copies 
of~\eqref{eq:comm_relations_so(1,4)}, the set of length scales 
$l(x)$ may form an arbitrary function of 
spacetime~\cite{Jennen:2014mba, Westman:2014yca}. Concomitantly, 
the cosmological constants~\eqref{eq:rel.Ll} of the local de 
Sitter spaces constitute a nonconstant cosmological function 
$\Lambda$.

Corresponding to the reductive nature of the 
relations~\eqref{eq:comm_relations_so(1,4)}, the Cartan 
connection is decomposed in a spin connection 
$\omega\ind{^a_{b\mu}}$ and vierbein $e\ind{^a_\mu}$, from which 
it follows that the vierbein has dimension of length.  
Furthermore, the curvature and torsion of the geometry are given 
by~\cite{Jennen:2014mba}
\begin{gather}
\label{eq:nonlin_curv_dSC}
  R\ind{^a_{b\mu\nu}} = B\ind{^a_{b\mu\nu}} + \frac{1}{l^2} 
  (e\ind{^a_\mu} e\ind{_{b\nu}} - e\ind{^a_\nu} e\ind{_{b\mu}})
  \\
\shortintertext{and}
\label{eq:nonlin_tors_dSC}
  T\ind{^a_{\mu\nu}} = G\ind{^a_{\mu\nu}} - \pd_\mu \ln l\, 
  e\ind{^a_\nu} + \pd_\nu \ln l\, e\ind{^a_\mu},
\end{gather}
where the two-forms
\begin{gather}
\label{eq:ext_cov_der_spin}
  B\ind{^a_{b\mu\nu}} = \pd_\mu \omega\ind{^a_{b\nu}} - \pd_\nu 
  \omega\ind{^a_{b\mu}} + \omega\ind{^a_{c\mu}} 
  \omega\ind{^c_{b\nu}} - \omega\ind{^a_{c\nu}} 
  \omega\ind{^c_{b\mu}}\\
\shortintertext{and}
\label{eq:ext_cov_der_vier}
  G\ind{^a_{\mu\nu}} = \pd_\mu e\ind{^a_\nu} - \pd_\nu 
  e\ind{^a_\mu} + \omega\ind{^a_{b\mu}} e\ind{^b_\nu} 
  - \omega\ind{^a_{b\nu}} e\ind{^b_\mu}
\end{gather}
are the exterior covariant derivatives of the spin connection, 
respectively, vierbein. When we solve~\eqref{eq:ext_cov_der_vier} 
for the spin connection, the Ricci theorem is recovered:
\begin{equation}
\label{eq:RicciTheor}
  \omega\ind{^a_{b\mu}} = \mring{\omega}\ind{^a_{b\mu}} 
  + K\ind{^a_{b\mu}},
\end{equation}
where $\mring{\omega}\ind{^a_{b\mu}} 
= \tfrac{1}{2}e\ind{^c_\mu}(\Omega\ind{_{bc}^a} 
+ \Omega\ind{_b^a_c} + \Omega\ind{_c^a_b})$ is the Levi-Civita 
spin connection, with $\Omega_{abc}  = e\ind{_a^\mu} 
e\ind{_b^\nu}(\pd_\mu e_{c\nu} - \pd_\nu e_{c\mu})$ the 
coefficients of anholonomy, and where
\begin{equation}
\label{eq:relContTors}
  K\ind{^a_{b\mu}} = \frac{1}{2} (G\ind{^a_{\mu b}} 
  + G\ind{_\mu^a_b} + G\ind{_b^a_\mu})
\end{equation}
is the contortion of $\omega\ind{^a_{b\mu}}$.

Subsequently, it is straightforward to define algebraic covariant 
differentiation by $D_\mu = \pd_\mu + \omega_\mu$, whereas 
$\nabla_\mu = \pd_\mu + \Gamma_\mu$ stands for the corresponding 
spacetime covariant derivative, i.e., $\Gamma\ind{^\rho_{\nu\mu}} 
= e\ind{_a^\rho} D_\mu e\ind{^a_\nu}$. A metric structure is 
readily constructed as well by defining $g_{\mu\nu} 
= e\ind{^a_\mu} e\ind{_{a\nu}}$, a symmetric tensor that is 
covariantly constant.

To conclude we remark that the pointwise decomposition of the 
Cartan connection in a spin connection and vierbein is manifestly 
local Lorentz invariant. The decomposition, albeit arbitrary, is 
not unique, for the possibilities are in a bijective 
correspondence with sections of the fibre bundle $Q[dS]$ of de 
Sitter spaces, associated with a principal $SO(1,4)$ bundle $Q$ 
over spacetime~\cite{husemoller:1966fibre}. The hidden 
symmetries, which are generic local de Sitter transformations 
that do not belong to the Lorentz subgroup, are manifestly 
restored when the geometric objects are constructed such that 
they have an explicit dependence on the section chosen.  This can 
be achieved through a nonlinear realization of the Cartan 
connection and it has been shown in~\cite{Jennen:2014mba} that 
such a nonlinear de Sitter--Cartan geometry with a cosmological 
function has a structure identical to the one outlined above.  
Therefore, we may assume the geometry of this section to be 
$SO(1,4)$ invariant.

\section{Cartan geometric structure of teleparallel gravity}
\label{sec:tg}

In order to make manifest that teleparallel gravity is 
constructed over a nonlinear Riemann--Cartan geometry, we briefly 
review its fundamentals~\cite{aldrovandi:2012tele}. These 
fundamentals were originally devised to form a gauge theory for 
the Poincar\'e translations. By pinning down its Cartan geometric 
structure, it will be easy to understand how to allow for 
a different kind of kinematics in Sec.~\ref{sec:dStg}.

Because kinematics is governed by the Poincar\'e group, spacetime 
$\mc{M}$ is the base manifold of a principal $ISO(1,3)$ bundle 
$Q$, while the local Minkowski spaces are the fibres $\{M_x \,|\, 
x \in \mc{M}\}$ of the associated bundle $Q \times_{ISO(1,3)}M$.  
The $\mf{iso}(1,3)$-valued connection $A_\mu dx^\mu$ determines 
how adjacent fibres compare under an infinitesimal displacement 
$dx^\mu$.

In the absence of gravity the connection is flat, and spacetime 
can be identified with every local Minkowski space as follows. We 
adopt spacetime coordinates $x^\mu = \delta^\mu_a \xi^a$ for the 
Cartesian system $\xi^a$, while the section $\xi^a(x) 
= \delta^a_\mu x^\mu$ of $Q[M]$ defines the points of tangency 
between the local Minkowski spaces and spacetime. It follows that 
the vierbein is given trivially by $e\ind{^a_\mu} 
= \delta^a_\mu$. After a general coordinate transformation 
$\delta^\mu_a \xi^a \mapsto x^\mu(\xi)$ the vierbein assumes the 
form $e\ind{^a_\mu} = \pd_\mu \xi^a$. There yet remains the 
freedom to consider local $ISO(1,3)$ transformations on the 
tangent Minkowski spaces. Since elements of $M_x$ transform as 
vectors under local Lorentz transformations $\xi^a \mapsto 
\Lambda\ind{^a_b} \xi^b$, we provide the vierbein with 
a corresponding connection, i.e., $e\ind{^a_\mu} = \pd_\mu\xi^a 
+ A\ind{^a_{b\mu}}\xi^b$. The connection is purely inertial, 
i.e., it takes the form $\Lambda\ind{^a_c} \pd_\mu 
\Lambda\ind{_b^c}$, so that no gravitational degrees of freedom 
are attributed to it.  Lastly, the components of Lorentz vectors 
are invariant under local Poincar\'e translations $\xi^a \mapsto 
\xi^a + \epsilon^a$.  Therefore, one includes a term in its 
definition
\begin{equation}
\label{eq:nonlin_vier_RC}
  e\ind{^a_\mu} = \pd_\mu\xi^a + A\ind{^a_{b\mu}}\xi^b 
  + A\ind{^a_\mu},
\end{equation}
which transforms as $A\ind{^a_\mu} \mapsto A\ind{^a_\mu} 
- \pd_\mu \epsilon^a - A\ind{^a_{b\mu}} \epsilon^b$. Note that 
the connection $A\ind{^a_{b\mu}}$ is invariant under local 
Poincar\'e translations, hence
\begin{equation}
\label{eq:nonlin_spin_RC}
  \omega\ind{^a_{b\mu}} = A\ind{^a_{b\mu}}
\end{equation}
is a well-defined spin connection, being independent of the gauge 
$\xi^a$ chosen.

The curvature and torsion then take the form
\begin{gather}
\label{eq:nonlin_curv_RC}
  B\ind{^a_{b\mu\nu}} = F\ind{^a_{b\mu\nu}}
  \\
  \shortintertext{and}
  \label{eq:nonlin_tors_RC}
  G\ind{^a_{\mu\nu}} = \xi^b F\ind{^a_{b\mu\nu}} 
  + F\ind{^a_{\mu\nu}},
\end{gather}
where the two-forms $F\ind{^a_{b\mu\nu}}$ and 
$F\ind{^a_{\mu\nu}}$ are the exterior covariant derivatives of 
$A\ind{^a_{b\nu}}$ and $A\ind{^a_\nu}$, respectively, with 
respect to $A\ind{^a_{b\mu}}$. Because the spin 
connection~\eqref{eq:nonlin_spin_RC} is purely inertial, the 
curvature~\eqref{eq:nonlin_curv_RC} vanishes, so that the torsion 
is determined entirely by $F\ind{^a_{\mu\nu}}$, which itself is 
nonzero only if $A\ind{^a_\mu} \neq \pd_\mu \epsilon^a 
+ A\ind{^a_{b\mu}} \epsilon^b$ for every choice of $\epsilon^a$, 
namely, it cannot be set to zero everywhere by a local Poincar\'e 
translation. The form $A\ind{^a_\mu}$ is therefore generally 
given the role of gravitational gauge potential, whereas the 
torsion is said to be its field strength.

The geometry defined by the spin 
connection~\eqref{eq:nonlin_spin_RC} and 
vierbein~\eqref{eq:nonlin_vier_RC}, together with their 
curvature~\eqref{eq:nonlin_curv_RC} and 
torsion~\eqref{eq:nonlin_tors_RC}, is a nonlinear Riemann--Cartan 
geometry~\cite{stelle.west:1980ds, Jennen:2014mba}. This 
signifies that Poincar\'e transformations acting on these objects 
are realized nonlinearly by elements of its Lorentz subgroup.  
Local Poincar\'e translations, for example, are realized 
trivially by the identity matrix. The implied invariance of the 
vierbein and torsion under such translations is a crucial 
ingredient to allow for the interpretation of teleparallel 
gravity as a gauge theory.  Indeed, as they play the role of 
covariant derivative and field strength, respectively, they must 
transform with the adjoint representation of the gauge group, 
which is the trivial representation due to the latter's abelian 
nature.

Because the set of de Sitter translations do not constitute 
a group, it does not appear feasible to construct teleparallel 
gravity with local kinematics regulated by $SO(1,4)$ through the 
gauge paradigm. By observing that the structure underlying 
teleparallel gravity is a nonlinear Riemann--Cartan geometry, it 
is natural to incorporate local de Sitter kinematics by 
generalizing for a nonlinear de Sitter--Cartan geometry.

\section{de Sitter teleparallel gravity}
\label{sec:dStg}

\subsection{Fundamentals of de Sitter teleparallel gravity}

Whilst the geometry summarized in Sec.~\ref{sec:dSCgeo} is the 
mathematical framework we shall employ to model teleparallel 
gravity with a cosmological function $\Lambda$, there remains to 
be specified how precisely it intends to accommodate the 
kinematics due to $\Lambda$ and the dynamical degrees of freedom 
of the gravitational field and the cosmological function.

To begin with, it is postulated that a gravitational field is 
present if and only if the exterior covariant 
derivative~\eqref{eq:ext_cov_der_vier} of the vierbein has 
a value not equal to zero. This characterization to indicate 
whether or not there are gravitational degrees of freedom is 
formally the same as in teleparallel gravity, which may be argued 
to be natural. \emph{A priori}, we would like to generalize for 
a different kinematics only and not alter the geometrical 
representation of the dynamics of the gravitational field.  
Nevertheless, this does not mean that the dynamics itself remains 
unaltered. On the contrary, we shall see in 
Sec.~\ref{ssec:dyn_grav_field} how the presence of a cosmological 
function modifies the gravitational field equations.

In further similarity with teleparallel gravity, the spin 
connection does not bear any gravitational degrees of freedom.  
Being a connection for local Lorentz transformations, it 
naturally continues to represent fictitious forces existing in 
a certain class of frames. The final and most important issue 
that must be settled in specifying for the geometry is then--- 
\emph{How are the local kinematics, whose defining group in the 
  presence of the cosmological function is $SO(1,4)$, accounted 
  for?}

This question is given an answer by postulating that the 
curvature~\eqref{eq:nonlin_curv_dSC} vanishes at every spacetime 
event, i.e.,
\begin{equation}
\label{eq:kine_curv_dStg}
  B\ind{^a_{b\mu\nu}} = -\frac{\Lambda}{3} (e\ind{^a_\mu} 
  e\ind{_{b\nu}} - e\ind{^a_\nu} e\ind{_{b\mu}}).
\end{equation}
The curvature of the spin connection hence equals the curvature 
of the Levi-Civita connection on a de Sitter space with 
cosmological constant given by $\Lambda$, which varies from point 
to point. If the cosmological function goes to zero over the 
whole of spacetime, the geometric structure of teleparallel 
gravity is recovered. The prescription~\eqref{eq:kine_curv_dStg} 
to implement the kinematics in the geometric framework is of 
great importance, for the kinematic effects will be observable as 
fictitious forces between adjacent free-falling particles, 
something which will be clarified in Sec.~\ref{ssec:part_mech}.

\subsection{Particle mechanics and kinematic effects}
\label{ssec:part_mech}

The motion of a particle of nonzero rest mass $m$ in the presence 
of a gravitational field and cosmological function is determined 
by the action $(c = 1)$
\begin{equation}
\label{eq:actionWLparticledSTG}
  \mc{S} = -m \int u_a e^a,
\end{equation}
where $u^a = e\ind{^a_\mu} dx^\mu/d\tau$ is the four-velocity of 
the particle. Hence, as usual~\eqref{eq:actionWLparticledSTG} is 
proportional to the particle's proper time $\tau$. The equations 
of motion are given by
\begin{equation}
\label{eq:partEOM_dStg}
  u^\rho D_\rho u^a = K\ind{^a_{b\rho}} u^b u^\rho,
\end{equation}
which are identical in form to the ones governing particle 
mechanics in teleparallel gravity~\cite{deAndrade:1997qt}, i.e., 
when the cosmological function vanishes. In 
particular,~\eqref{eq:partEOM_dStg} complies with the weak 
equivalence principle, yet a breakdown of the latter most likely 
could be coped with along the lines it is done in teleparallel 
gravity~\cite{Aldrovandi:2003pa}. Despite the fact it is not 
immediately obvious from~\eqref{eq:partEOM_dStg}, a nonzero 
cosmological function has an impact on the motion of particles.  
The first change is rather indirect and stems from a modification 
in the gravitational field equations, thus altering the value of 
the contortion for a given distribution of energy-momentum that 
sources gravity. 

The second change reflects the alteration in kinematics, now 
regulated by the de Sitter group. In order to clarify this, we 
consider a one-parameter family of solutions $x_\sigma(\tau)$ 
of~\eqref{eq:partEOM_dStg}, pa\-ram\-e\-trized by $\sigma$.  
These solutions form a two-dimensional surface, to which the 
vector fields $u = d/d\tau$ and $v = d/d\sigma$ are tangent. For 
every $\sigma$, $u$ is the four-velocity of the particle with 
world line $x_\sigma$. The field $v$ is tangent to constant 
$\tau$ slices, connecting the world lines of neighboring 
particles during their motion through spacetime. The vector 
field~\cite{carroll:sg}
\begin{equation*}
  \mf{a}^a = u^\mu D_\mu (u^\nu D_\nu v^a)
\end{equation*}
is therefore the relative acceleration between the world lines, 
measured by a free-falling observer. Because $[u,v]$ equals zero 
and
\begin{equation*}
u^\mu D_\mu v^a - v^\mu D_\mu u^a = u^\mu v^\nu 
G\ind{^a_{\mu\nu}},
\end{equation*}
while $u^a$ satisfies~\eqref{eq:partEOM_dStg}, the relative 
acceleration may be written as
\begin{equation}
\label{eq:rel_acc_dStg}
  \mf{a}^a = u^\mu v^\nu u^b B\ind{^a_{b\mu\nu}}
  + v^\mu D_\mu (K\ind{^a_{b\nu}} u^b u^\nu) + u^\mu D_\mu 
  (u^\lambda v^\nu G\ind{^a_{\lambda\nu}}),
\end{equation}
where $B\ind{^a_{b\mu\nu}}$ is the
curvature~\eqref{eq:kine_curv_dStg} of the local de Sitter 
spaces. The first term is therefore present only when the 
cosmological function is nonzero.

Equation~\eqref{eq:rel_acc_dStg} is a chief result of de Sitter 
teleparallel gravity, for it describes what the phenomenology is 
of the local de Sitter kinematics. The last two terms are 
dynamical in nature and come from a nonhomogeneous gravitational 
field. The first term originates in the cosmological function 
$\Lambda$, as can be seen from~\eqref{eq:kine_curv_dStg}, and is 
caused by the kinematics.  This contribution manifests itself in 
that two particles separated by the infinitesimal $v^a$ deviate 
as if they were moving in a de Sitter space with cosmological 
constant $\Lambda$. Hence, two neighboring free-falling particles 
have world lines that deviate, not only because they move in 
a nonhomogeneous gravitational field, but also because of the 
kinematics that is determined by the cosmological function.  
According to this approach, dark energy has its origins in the 
cosmological function and reveals itself as a kinematic effect.

\subsection{Dynamics of the gravitational field and the 
  cosmological function}
\label{ssec:dyn_grav_field}

Having specified the particle mechanics caused by a gravitational 
field and cosmological function, we now prescribe the dynamics of 
the latter two themselves. We shall define the gravitational 
action as a reasonable generalization of the action for the 
gravitational field in teleparallel gravity, for which the 
Lagrangian is given by~\cite{Maluf:2013gaa}
\begin{equation*}
  \mc{L}_\mrm{tg} = \frac{1}{4} G \ind{^\rho_{\mu\nu}} 
  G\ind{_\rho^{\mu\nu}} + \frac{1}{2} G\ind{^\rho_{\mu\nu}} 
  G\ind{^{\nu\mu}_\rho} - G\ind{^\nu_{\mu\nu}} 
  G\ind{^{\rho\mu}_\rho}.
\end{equation*}
In teleparallel gravity, the two-form $G\ind{^a_{\mu\nu}}$ is the 
torsion of the underlying Riemann--Cartan geometry. Because we 
have generalized for a de Sitter--Cartan geometry, for which the 
torsion is given by~\eqref{eq:nonlin_tors_dSC}, it appears 
a natural proposition to define the Lagrangian for de Sitter 
teleparallel gravity as
\begin{equation*}
  \mc{L}_\mrm{dStg} = \frac{1}{4} T \ind{^\rho_{\mu\nu}} 
  T\ind{_\rho^{\mu\nu}} + \frac{1}{2} T\ind{^\rho_{\mu\nu}} 
  T\ind{^{\nu\mu}_\rho} - T\ind{^\nu_{\mu\nu}} 
  T\ind{^{\rho\mu}_\rho}.
\end{equation*}
Substituting for~\eqref{eq:nonlin_tors_dSC} restates this 
function as a Lagrangian for the gravitational field and 
cosmological function, namely,
\begin{equation}
\label{eq:lagrangian_dStg}
  \mc{L}_\mrm{dStg} = \mc{L}_\mrm{tg} - \frac{3}{2} \pd_\mu \ln 
  \Lambda\, \pd^\mu \ln \Lambda
  - 2 G\ind{^{\mu\nu}_\mu} \pd_\nu \ln \Lambda.
\end{equation}
The action for de Sitter teleparallel gravity is thus given by 
$(c = \hbar = 1)$
\begin{equation}
\label{eq:action_dStg}
  \mc{S}_\mrm{dStg} = \frac{1}{2\kappa} \int d^4 x \, e \, 
  \mc{L}_\mrm{dStg},
\end{equation}
where $\kappa = 8\pi G_N$ and $e = \det e\ind{^a_\mu}$.

The action~\eqref{eq:action_dStg} reminds, on the one hand, of 
the scheme in which scalar-tensor theories modify gravity in the 
framework of general relativity~\cite{Brans:1961sx, Dicke:1961gz, 
  Bergmann:1968ve, Sotiriou:2008rp, Tsujikawa:2010zza}, or, on 
the other hand, of teleparallel dark energy, where a scalar field 
is coupled nonminimally to teleparallel 
gravity~\cite{Geng:2011aj, Geng:2011ka, Xu:2012jf}. To be 
precise, it specifies for a gravitational sector modeled by 
teleparallel gravity--- for a spin connection with 
curvature~\eqref{eq:kine_curv_dStg}--- that interacts with the 
cosmological function due to a nonminimal coupling between the 
trace of the exterior covariant derivative of the vierbein and 
the logarithmic derivative of $\Lambda$. A theory quite similar 
in structure was discussed in~\cite{Otalora:2014aoa}. Despite the 
similarity, there is a crucial discrepancy it has in common with 
any of the other modifications of general relativity or 
teleparallel gravity that introduce nonminimal couplings to 
scalar fields. Usually, these fields are added to the theory in 
a manner rather reminiscent of \emph{ad hoc} hypotheses, and are 
not an essential feature of the spacetime geometry. In de Sitter 
teleparallel gravity by contrast, the scalar field is the 
cosmological function, which forms an integral part of the 
geometric structure and, moreover, quantifies the kinematics 
locally governed by the de Sitter group in the sense of 
Sec.~\ref{ssec:part_mech}. 

Note that the cosmological function appears in the action only 
through its logarithmic derivative, which is a direct consequence 
of~\eqref{eq:nonlin_tors_dSC}. Factors of $\pd_\mu \ln \Lambda$ 
are naturally dimensionless, which renders a correct overall 
dimension for the action. Because spacetime coordinates are 
numbers and the dimension of the vierbein components is that of 
length, i.e., $[e\ind{^a_\mu}] = L$, while $[\kappa] = L^2$, $[e] 
= L^4$, $[g^{\mu\nu}] = L^{-2}$, and $[G\ind{^{\rho\mu}_\rho}] 
= L^{-2}$, the action~\eqref{eq:action_dStg} indeed has dimension 
of $\hbar = 1$.

The field equations for the vierbein are
\begin{multline}
\label{eq:fieldEqs_dStg_expl}
   D_\rho \big(e\, W\ind{_a^{\rho\mu}}\big) + e\, t\ind{_a^\mu} 
   - 2e\, G\ind{^{\rho\mu}_\rho} e\ind{_a^\nu} \pd_\nu \ln 
   \Lambda \\ - 3 e\, e\ind{_a^\rho} \pd_\rho \ln \Lambda\, 
   \pd^\mu \ln \Lambda
   + \frac{3}{2} e\, e\ind{_a^\mu} \pd_\rho \ln \Lambda\, 
   \pd^\rho \ln \Lambda \hspace{1cm}
   \\
   - 2e\, e\ind{_a^\mu} \Box \ln \Lambda + 2e\, e\ind{_a^\rho} 
   \nabla_\rho \pd^\mu \ln \Lambda = 0,
\end{multline}
where $\Box\hspace{0.1em} = g^{\mu\nu} \nabla_\mu \pd_\nu$ is the 
d'Alembertian, while
\begin{equation*}
  W\ind{_a^{\mu\nu}} \equiv G\ind{_a^{\mu\nu}} 
  + G\ind{^{\nu\mu}_a} - G\ind{^{\mu\nu}_a} \\
  - 2e\ind{_a^\nu} G\ind{^{\lambda\mu}_\lambda} + 2e\ind{_a^\mu} 
  G\ind{^{\lambda\nu}_\lambda},
\end{equation*}
and
\begin{equation*}
  t\ind{_a^\mu} = G\ind{^b_{\rho a}} W\ind{_b^{\rho\mu}} 
  - e\ind{_a^\mu} \mathcal{L}_\mrm{tg}.
\end{equation*}
are the superpotential, respectively, the gravitational 
energy-momentum current~\cite{Maluf:2013gaa}. The gravitational 
field equations~\eqref{eq:fieldEqs_dStg_expl} solve for the 
components of the vierbein, but do not determine the cosmological 
function. In de Sitter teleparallel gravity $\Lambda$ is given 
its own dynamics, which is dictated by the field equation
\begin{equation}
\label{eq:field_eq_lambda}
  \Box \ln \Lambda + G\ind{^{\mu\rho}_\mu} \pd_\rho \ln \Lambda =
  -\frac{2}{3} ( \nabla_\mu G\ind{^{\rho\mu}_\rho} 
  + G\ind{^\mu_{\rho\mu}} G\ind{^{\nu\rho}_\nu} ).
\end{equation}

The coupling of matter fields to the gravitational sector is 
carried out by taking the sum of the matter action
\begin{equation*}
  \mc{S}_\mrm{m} = \int d^4x \, e \, \mc{L}_\mrm{m}
\end{equation*}
and the action~\eqref{eq:action_dStg} for the gravitational field 
and cosmological function. The energy-momentum current $\delta (e 
\mc{L}_\mrm{m})/\delta e\ind{^a_\mu}$ of matter is a source for 
the gravitational field equations~\eqref{eq:fieldEqs_dStg_expl}, 
but does not appear in the equation of 
motion~\eqref{eq:field_eq_lambda} for the cosmological function.  
According to this scheme, energy-momentum generates gravity, 
which in turn sources the cosmological function.

\section{Conclusions}
\label{sec:concl}

In the present work we formulated a theory of gravity consistent 
with local spacetime kinematics regulated by the de Sitter group, 
namely, de Sitter teleparallel gravity. It was made plain first 
that teleparallel gravity, a theory physically equivalent to 
general relativity, has the mathematical structure of a nonlinear 
Riemann--Cartan geometry. This inspired us to generalize for de 
Sitter kinematics by considering de Sitter--Cartan geometry in 
the presence of a nonconstant cosmological function~$\Lambda$. 

The theory consists of a gravitational sector described by 
teleparallel gravity that couples nonminimally to the 
cosmological function.  Dynamical degrees of freedom of the 
gravitational field are present if and only if the exterior 
covariant derivative of the vierbein does not vanish.  Further, 
the cosmological function has its own dynamics, sourced by the 
trace of the exterior covariant derivative of the vierbein, but 
not directly by the matter energy-momentum current.  It is thence 
similar in form to teleparallel dark energy, or scalar-tensor 
theories in the framework of general relativity. Because of this 
similarity, the analysis regarding the phenomenology of dark 
energy within some of these theories is expected to hold to 
a high degree within the framework of de Sitter teleparallel 
gravity.

On the other hand, a crucial difference between these models and 
the theory here proposed is that the cosmological function 
modifies the local kinematics of spacetime. Indeed, at every 
spacetime point we put forward that the curvature of the spin 
connection is equal to the curvature of the Levi-Civita 
connection of a de Sitter space with cosmological constant given 
by the value of the cosmological function. We saw that such 
a choice gives rise to a kinematic contribution in the deviation 
equation for the world lines of adjacent free-falling particles, 
that is, they undergo a relative acceleration that is kinematic 
in origin. This result is arguably the one of most importance of 
this article, for it specifies in exactly what manner the 
kinematics due to the cosmological function are to be observed.  
Hence, dark energy may be interpreted as a kinematic effect or, 
alternatively, as the cosmological function causing this effect.

It is interesting to note that there exists a link between the 
dynamics and kinematics of the theory, in the sense that the 
value of the cosmological function is determined dynamically by 
its interaction with the gravitational field, while the resulting 
value determines the local spacetime kinematics, which in its 
turn affects the motion of matter. The theory thus gives 
a precise model that prescribes how the kinematics of high energy 
physics may be modified locally and becomes 
spacetime-dependent~\cite{Mansouri:2002cg}. Although there is 
a connection between them, dynamics and kinematics remain 
logically separated in the geometric representation of de Sitter 
teleparallel gravity. Nontrivial dynamics gives way to the 
torsion of the de Sitter--Cartan geometry being nonzero, whereas 
the value of the curvature of the spin connection encodes the 
inertial effects of a given frame and the local de Sitter 
kinematics. This is a natural generalization of the geometric 
representation of teleparallel gravity, which is recovered when 
the cosmological function vanishes.

The new paradigm here presented to replace the Poincar\'e group 
with the de Sitter group as the set of transformations that 
govern local kinematics is further motivated by prospective 
problems to conciliate special relativity with the existence of 
an invariant length parameter at the Planck scale. Namely, the 
concomitant replacement of special relativity with an $SO(1,4)$ 
invariant special relativity allows for the existence of an 
invariant length parameter proportional to $\Lambda^{-2}$, while 
preserving the constancy of the speed of 
light~\cite{Aldrovandi:2006vr}.  Because the cosmological 
function is not restricted to be constant, its value can evolve 
with cosmological time, this model may be suitable to describe 
the evolution of the universe, which requires different values of 
the cosmological term at different epochs. For example, a huge 
cosmological term could drive inflation at the primordial 
universe. Afterwards, the cosmological term should decay to 
a small value in order to allow the formation of the structures 
we see today. Then, to account for the late--time acceleration in 
the universe expansion rate, the value of the cosmological term 
should somehow increase~\cite{Araujo:2015oqa}.

It goes without saying that the ideas implemented in this paper 
constitute just a new scenario for studying cosmology,
and further research needs to be conducted. For example, the 
field equations~\eqref{eq:fieldEqs_dStg_expl} 
and~\eqref{eq:field_eq_lambda} must be solved for a homogeneous 
and isotropic universe in order to find the time evolution of the 
scale factor and the cosmological function. Whether the dynamical 
model here presented is consistent with observed dark energy 
behavior may be verified in a model-independent way, i.e.,  
cosmographically~\cite{Visser:2004bf}, by comparing the 
calculated values for low order derivatives of the present-day 
scale factor with their observed values. Such cosmographic 
requirements previously have been applied to put constraints on 
the functional form of modified gravity Lagrangians, see, 
e.g.,~\cite{Capozziello:2015rda}. Another interesting question to 
be answered is how the kinematic contribution in the deviation 
equation~\eqref{eq:rel_acc_dStg} will leave its trace in the 
Raychaudhuri equation and what will be the implication thereof 
for the motion of particles in a gravitational field.

Finally should it be noted that if the dynamical 
model~\eqref{eq:lagrangian_dStg} were to be falsified by 
cosmographic observations, one would naturally be forced to find 
alternative Lagrangians for the gravitational field and 
cosmological function, but the paradigm of representing dark 
energy in the kinematics of spacetime in tele\-parallel gravity 
through the prescription~\eqref{eq:kine_curv_dStg} would remain 
a valid track to be considered further.

\section*{Acknowledgments}

The authors gratefully acknowledge financial support by CAPES, 
CNPq and FAPESP.

\section*{References}

\bibliographystyle{elsarticle-num}
\bibliography{arx.dStg.v2.bbl}

\begin{thebibliography}{10}
\expandafter\ifx\csname url\endcsname\relax
  \def\url#1{\texttt{#1}}\fi
\expandafter\ifx\csname urlprefix\endcsname\relax\def\urlprefix{URL }\fi
\expandafter\ifx\csname href\endcsname\relax
  \def\href#1#2{#2} \def\path#1{#1}\fi

\bibitem{aldrovandi:2012tele}
R.~Aldrovandi, J.~G. Pereira, Teleparallel Gravity: An Introduction, Springer,
  Dordrecht, 2012.

\bibitem{Aldrovandi:2003pa}
R.~Aldrovandi, J.~G. Pereira, K.~H. Vu, Gravitation without the equivalence
  principle, Gen. Rel. Grav. 36 (2004) 101--110.
\newblock \href {http://arxiv.org/abs/gr-qc/0304106}
  {\path{arXiv:gr-qc/0304106}}, \href
  {http://dx.doi.org/10.1023/B:GERG.0000006696.98824.4d}
  {\path{doi:10.1023/B:GERG.0000006696.98824.4d}}.

\bibitem{Peebles:2003cc}
P.~J.~E. Peebles, B.~Ratra, The cosmological constant and dark energy, Rev.
  Mod. Phys. 75 (2003) 559--606.
\newblock \href {http://arxiv.org/abs/astro-ph/0207347}
  {\path{arXiv:astro-ph/0207347}}, \href
  {http://dx.doi.org/10.1103/RevModPhys.75.559}
  {\path{doi:10.1103/RevModPhys.75.559}}.

\bibitem{Weinberg:2008bc}
S.~Weinberg, Cosmology, Oxford University Press, 2008.

\bibitem{Aldrovandi:2006vr}
R.~Aldrovandi, J.~P. Beltran~Almeida, J.~G. Pereira, {d}e {S}itter special
  relativity, Class. Quantum Grav. 24 (2007) 1385--1404.
\newblock \href {http://arxiv.org/abs/gr-qc/0606122}
  {\path{arXiv:gr-qc/0606122}}, \href
  {http://dx.doi.org/10.1088/0264-9381/24/6/002}
  {\path{doi:10.1088/0264-9381/24/6/002}}.

\bibitem{Wise:2010sm}
D.~K. Wise, {M}ac{D}owell-{M}ansouri gravity and {C}artan geometry, Class.
  Quantum Grav. 27 (2010) 155010.
\newblock \href {http://arxiv.org/abs/gr-qc/0611154}
  {\path{arXiv:gr-qc/0611154}}, \href
  {http://dx.doi.org/10.1088/0264-9381/27/15/155010}
  {\path{doi:10.1088/0264-9381/27/15/155010}}.

\bibitem{Jennen:2014mba}
H.~Jennen, {Cartan geometry of spacetimes with a nonconstant cosmological
  function $\Lambda$}, Phys. Rev. D90~(8) (2014) 084046.
\newblock \href {http://arxiv.org/abs/1406.2621} {\path{arXiv:1406.2621}},
  \href {http://dx.doi.org/10.1103/PhysRevD.90.084046}
  {\path{doi:10.1103/PhysRevD.90.084046}}.

\bibitem{Alekseevsky:1995cc}
D.~V. Alekseevsky, P.~W. Michor, Differential geometry of {C}artan connections,
  Publ. Math. Debrecen 47 (1995) 349--375.
\newblock \href {http://arxiv.org/abs/math/9412232}
  {\path{arXiv:math/9412232}}.

\bibitem{sharpe1997diff_geo}
R.~W. Sharpe, Differential Geometry: {C}artan's Generalization of {K}lein's
  {E}rlangen Program, Springer, New York, 1997.

\bibitem{Wise:2009fu}
D.~K. Wise, Symmetric space {C}artan connections and gravity in three and four
  dimensions, SIGMA 5 (2009) 080.
\newblock \href {http://arxiv.org/abs/0904.1738} {\path{arXiv:0904.1738}},
  \href {http://dx.doi.org/10.3842/SIGMA.2009.080}
  {\path{doi:10.3842/SIGMA.2009.080}}.

\bibitem{Westman:2014yca}
H.~F. Westman, T.~G. Zlosnik, An introduction to the physics of {C}artan
  gravity, Ann. Phys. 361 (2015) 330--376.
\newblock \href {http://arxiv.org/abs/1411.1679} {\path{arXiv:1411.1679}},
  \href {http://dx.doi.org/10.1016/j.aop.2015.06.013}
  {\path{doi:10.1016/j.aop.2015.06.013}}.

\bibitem{husemoller:1966fibre}
D.~Husem{\"o}ller, Fibre Bundles, Springer, New York, 1966.

\bibitem{stelle.west:1980ds}
K.~S. Stelle, P.~C. West, Spontaneously broken de~{S}itter symmetry and the
  gravitational holonomy group, Phys. Rev. D21 (1980) 1466--1488.
\newblock \href {http://dx.doi.org/10.1103/PhysRevD.21.1466}
  {\path{doi:10.1103/PhysRevD.21.1466}}.

\bibitem{deAndrade:1997qt}
V.~C. de~Andrade, J.~G. Pereira, Gravitational lorentz force and the
  description of the gravitational interaction, Phys. Rev. D56 (1997)
  4689--4695.
\newblock \href {http://arxiv.org/abs/gr-qc/9703059}
  {\path{arXiv:gr-qc/9703059}}, \href
  {http://dx.doi.org/10.1103/PhysRevD.56.4689}
  {\path{doi:10.1103/PhysRevD.56.4689}}.

\bibitem{carroll:sg}
S.~Carroll, Spacetime and Geometry: an Introduction to General Relativity,
  Addison Wesley, San Francisco, 2004.

\bibitem{Maluf:2013gaa}
J.~W. Maluf, The teleparallel equivalent of general relativity, Ann. Phys.
  (Berlin) 525 (2013) 339--357.
\newblock \href {http://arxiv.org/abs/1303.3897} {\path{arXiv:1303.3897}},
  \href {http://dx.doi.org/10.1002/andp.201200272}
  {\path{doi:10.1002/andp.201200272}}.

\bibitem{Brans:1961sx}
C.~Brans, R.~H. Dicke, {M}ach's principle and a relativistic theory of
  gravitation, Phys. Rev. 124 (1961) 925--935.
\newblock \href {http://dx.doi.org/10.1103/PhysRev.124.925}
  {\path{doi:10.1103/PhysRev.124.925}}.

\bibitem{Dicke:1961gz}
R.~H. Dicke, {M}ach's principle and invariance under transformation of units,
  Phys. Rev. 125 (1962) 2163--2167.
\newblock \href {http://dx.doi.org/10.1103/PhysRev.125.2163}
  {\path{doi:10.1103/PhysRev.125.2163}}.

\bibitem{Bergmann:1968ve}
P.~G. Bergmann, Comments on the scalar tensor theory, Int. J. Theor. Phys. 1
  (1968) 25--36.
\newblock \href {http://dx.doi.org/10.1007/BF00668828}
  {\path{doi:10.1007/BF00668828}}.

\bibitem{Sotiriou:2008rp}
T.~P. Sotiriou, V.~Faraoni, $f(r)$ theories of gravity, Rev. Mod. Phys. 82
  (2010) 451--497.
\newblock \href {http://arxiv.org/abs/0805.1726} {\path{arXiv:0805.1726}},
  \href {http://dx.doi.org/10.1103/RevModPhys.82.451}
  {\path{doi:10.1103/RevModPhys.82.451}}.

\bibitem{Tsujikawa:2010zza}
S.~Tsujikawa, Modified gravity models of dark energy, Lect. Notes Phys. 800
  (2010) 99--145.
\newblock \href {http://arxiv.org/abs/1101.0191} {\path{arXiv:1101.0191}},
  \href {http://dx.doi.org/10.1007/978-3-642-10598-2_3}
  {\path{doi:10.1007/978-3-642-10598-2_3}}.

\bibitem{Geng:2011aj}
C.-Q. Geng, C.-C. Lee, E.~N. Saridakis, Y.-P. Wu, 'teleparallel' dark energy,
  Phys. Lett. B704 (2011) 384--387.
\newblock \href {http://arxiv.org/abs/1109.1092} {\path{arXiv:1109.1092}},
  \href {http://dx.doi.org/10.1016/j.physletb.2011.09.082}
  {\path{doi:10.1016/j.physletb.2011.09.082}}.

\bibitem{Geng:2011ka}
C.-Q. Geng, C.-C. Lee, E.~N. Saridakis, Observational constraints on
  teleparallel dark energy, JCAP 1201 (2012) 002.
\newblock \href {http://arxiv.org/abs/1110.0913} {\path{arXiv:1110.0913}},
  \href {http://dx.doi.org/10.1088/1475-7516/2012/01/002}
  {\path{doi:10.1088/1475-7516/2012/01/002}}.

\bibitem{Xu:2012jf}
C.~Xu, E.~N. Saridakis, G.~Leon, Phase-space analysis of teleparallel dark
  energy, JCAP 1207 (2012) 005.
\newblock \href {http://arxiv.org/abs/1202.3781} {\path{arXiv:1202.3781}},
  \href {http://dx.doi.org/10.1088/1475-7516/2012/07/005}
  {\path{doi:10.1088/1475-7516/2012/07/005}}.

\bibitem{Otalora:2014aoa}
G.~Otalora, A novel teleparallel dark energy model\href
  {http://arxiv.org/abs/1402.2256} {\path{arXiv:1402.2256}}.

\bibitem{Mansouri:2002cg}
F.~Mansouri, Nonvanishing cosmological constant $\lambda$, phase transitions,
  and $\lambda$-dependence of high-energy processes, Phys. Lett. B538 (2002)
  239--245.
\newblock \href {http://arxiv.org/abs/hep-th/0203150}
  {\path{arXiv:hep-th/0203150}}, \href
  {http://dx.doi.org/10.1016/S0370-2693(02)02022-1}
  {\path{doi:10.1016/S0370-2693(02)02022-1}}.

\bibitem{Araujo:2015oqa}
A.~Araujo, H.~Jennen, J.~G. Pereira, A.~C. Sampson, L.~L. Savi, {On the
  spacetime connecting two aeons in conformal cyclic cosmology}, Gen. Rel.
  Grav. 47~(12) (2015) 151.
\newblock \href {http://arxiv.org/abs/1503.05005} {\path{arXiv:1503.05005}},
  \href {http://dx.doi.org/10.1007/s10714-015-1991-4}
  {\path{doi:10.1007/s10714-015-1991-4}}.

\bibitem{Visser:2004bf}
M.~Visser, {Cosmography: Cosmology without the Einstein equations}, Gen. Rel.
  Grav. 37 (2005) 1541--1548.
\newblock \href {http://arxiv.org/abs/gr-qc/0411131}
  {\path{arXiv:gr-qc/0411131}}, \href
  {http://dx.doi.org/10.1007/s10714-005-0134-8}
  {\path{doi:10.1007/s10714-005-0134-8}}.

\bibitem{Capozziello:2015rda}
S.~Capozziello, O.~Luongo, E.~N. Saridakis, {Transition redshift in $f(T)$
  cosmology and observational constraints}, Phys. Rev. D91~(12) (2015) 124037.
\newblock \href {http://arxiv.org/abs/1503.02832} {\path{arXiv:1503.02832}},
  \href {http://dx.doi.org/10.1103/PhysRevD.91.124037}
  {\path{doi:10.1103/PhysRevD.91.124037}}.

\end{thebibliography}

\end{document}